\documentclass{appolb}
\usepackage{graphicx}

\usepackage{hyperref}
\usepackage{mathrsfs}
\usepackage{amsmath,amssymb,dsfont,placeins}
\usepackage{verbatim}
\usepackage{stmaryrd,xcolor}
\usepackage{amsfonts,graphicx,epsfig,float,ulem}
\usepackage{bbm}

\begin{document}
 
\eqsec  

\title{Neutrino mass spectrum from the seesaw extension
\thanks{Presented at International Symposium on Multiparticle Dynamics, Kielce 2012}%
}
\author{Darius Jurciukonis, Thomas Gajdosik, Andrius Juodagalvis and Tomas Sabonis
\address{Vilnius University, Universiteto 3, LT-01513, Vilnius, Lithuania}
}

\maketitle
\begin{abstract}
The Standard Model includes neutrinos as massless particles, but
neutrino oscillations showed that neutrinos are not massless. A simple
extension of adding gauge singlet fermions to the particle spectrum
allows normal Yukawa mass terms for neutrinos. The smallness of the
neutrino masses can be well understood within the seesaw mechanism. We
analyse two cases of the minimal extension of the standard model when
one or two right-handed fields are added to the three left-handed
fields. A second Higgs doublet is included in our model. We calculate
the one-loop radiative corrections to the mass parameters which
produce mass terms for the neutral leptons. In both cases we
numerically analyse light neutrino masses as functions of the heavy
neutrinos masses. Parameters of the model are varied to find light
neutrino masses that are compatible with experimental data of solar
$\Delta m^2_\odot$ and atmospheric $\Delta m^2_\mathrm{atm}$ neutrino
oscillations for normal and inverted hierarchy.
\end{abstract}

\section{The model}

We extend the Standard Model (SM) by adding a second Higgs doublet and 
right-handed neutrino fields. The Yukawa Lagrangian
of the leptons is expressed by
\begin{equation}
\mathcal{L}_\mathrm{Y} = - \sum_{k=1}^{2}\, 
\left( \Phi_k^\dagger \bar \ell_R \Gamma_k
+ \tilde \Phi_k^\dagger \bar \nu_R \Delta_k \right) D_L
+ \mathrm{H.c.}
\label{Yukawa}
\end{equation}
in a vector and matrix notation, where $\tilde \Phi_k = i \tau_2
\Phi_k^\ast$.  In expression (\ref{Yukawa}) $\ell_R$, $\nu_R$, and
$D_L=(\nu_L\ \ell_L)^T$ are the vectors of the right-handed charged
leptons, of the right-handed neutrino singlets, and of the left-handed
lepton doublets, respectively, and $\Phi_k, k=1,2$ are the 
two Higgs doublets. The Yukawa coupling matrices $\Gamma_k$
are $n_L \times n_L$, while the $\Delta_k$ are $n_R \times n_L$.

In this model, spontaneous symmetry breaking of the SM gauge group is
achieved by the vacuum expectation values
$\langle\Phi_k\rangle_{\mathrm{vac}} =
\left( 0 , v_k/\sqrt{2}  \right)^T$, $k=1, 2$. By a unitary
rotation of the Higgs doublets we can achieve
$\langle\Phi^0_1\rangle_{\mathrm{vac}} = v/\sqrt{2} > 0$ and
$\langle\Phi^0_2\rangle_{\mathrm{vac}} = 0$ with $v \simeq 246$~GeV.
The charged-lepton mass matrix $M_\ell$ and the Dirac neutrino mass
matrix $M_D$ are
\begin{equation}
\hat{M}_\ell = \frac{ v}{\sqrt{2}}\, \Gamma_1
\quad {\rm and} \quad
M_D = \frac{v}{\sqrt{2}}\, \Delta_1\, ,
\label{M_ell,M_D}
\end{equation}
with the assumption that $\hat{M}_\ell = {\rm diag} \left( m_e,
m_\mu, m_\tau \right)$. The hat indicates that $\hat{M}_l$ is a diagonal matrix. The mass terms for the neutrinos can be
written in a compact form with a $(n_L+n_R) \times (n_L+n_R)$
symmetric mass matrix
\begin{equation}\label{Mneutr}
M_{\nu} = 
\left(\renewcommand{\arraystretch}{0.8} \begin{array}{cc} 0 & M_D^T \\
M_D & \hat{M}_R \end{array} \right).
\end{equation}
$M_{\nu}$ can be diagonalized as
\begin{equation}\label{Mtotal}
U^T M_{\nu}\, U = \hat m
= \mathrm{diag} \left( m_1, m_2, \ldots, m_{n_L+n_R} \right),
\end{equation}
where the $m_i$ are real and non-negative.  In order to implement the
seesaw mechanism \cite{seesaw,Schechter} we assume that the elements
of $M_D$ are of order $m_D$ and those of $M_R$ are of order $m_R$,
with $m_D \ll m_R$.  Then the neutrino masses $m_i$ with $i=1, 2,
\ldots, n_L$ are of order $m_D^2/m_R$, while those with $i = n_L+1,
\ldots, n_L+n_R$ are of order $m_R$.  It is useful to decompose the
$(n_L+n_R) \times (n_L+n_R)$ unitary matrix $U$ as $U =
\left(U_L , U_R^\ast \right)^T$, where the submatrix
$U_L$ is $n_L \times (n_L+n_R)$ and the submatrix $U_R$ is $n_R \times
(n_L+n_R)$ \cite{GN89,GL02}.  With these submatrices, the left- and
right-handed neutrinos are written as linear superpositions of the
$n_L+n_R$ physical Majorana neutrino fields $\chi_i$: $\nu_L = U_L P_L
\chi$ and $\nu_R = U_R P_R \chi$, where $P_L$ and $P_R$ are the
projectors of chirality.

It is possible to express the couplings of the model in terms of the mass
eigenfields, where three neutral particles namely, the $Z$ boson, the neutral Goldstone boson $G^0$ and the Higgs bosons $H^0_b$ couple to neutrinos. The full formalism for the scalar sector of the multi-Higgs-doublet SM is
given in Refs.~\cite{GN89,GL02}.

Once the one-loop corrections are taken into account
the neutral fermion mass matrix is given by
\begin{equation}\label{M1}
M^{(1)}_\nu = \left(\renewcommand{\arraystretch}{0.8} \begin{array}{cc}
\delta M_L & M_D^T+\delta M_D^T \\
M_D+\delta M_D    & \hat{M}_R+\delta M_R\end{array} \right)\approx
\left(\renewcommand{\arraystretch}{0.8} \begin{array}{cc}
\delta M_L & M_D^T \\
M_D   & \hat{M}_R\end{array} \right),
\end{equation}
where the $0_{3\times3}$ matrix appearing at tree level (\ref{Mneutr})
is replaced by the contribution $\delta M_L$. This correction is a
symmetric matrix, it dominates among all the sub-matrices of corrections.
Neglecting the sub-dominant pieces in (\ref{M1}), one-loop corrections to the neutrino masses originate via the self-energy function $\Sigma_L^S (0)=\Sigma_L^{S(Z)}(0)+\Sigma_L^{S(G^0)}(0)+\Sigma_L^{S(H^0)}(0)$, where the $\Sigma_L^{S(Z,G^0,H^0)}(0)$ functions arise from the
self-energy Feynman diagrams and are evaluated at zero external momentum squared.
Each diagram contains a divergent
piece but when summing up the three contributions the result turns out
to be finite.

The final expression for one-loop corrections is given by \cite{Grimus:2002nk} 
\begin{eqnarray}
\delta M_L &=&
\sum_{b} \frac{1}{32 \pi^2}\, \Delta_b^T U_R^\ast \hat m
\left(
    \frac{\hat {m}^2}{m_{H^0_b}^2}-\mathbbm{1}
  \right)^{-1}\hspace{-5pt}
  \ln\left(\frac{\hat {m}^2}{m_{H^0_b}^2}\right) U_R^\dagger \Delta_b \notag \\
&&+ \frac{3 g^2}{64 \pi^2 m_W^2}\, M_D^T U_R^\ast \hat m
\left(
    \frac{\hat {m}^2}{m_Z^2}-\mathbbm{1}
  \right)^{-1}\hspace{-5pt}
  \ln\left(\frac{\hat {m}^2}{m_Z^2}\right) U_R^\dagger M_D,
\label{corrections}
\end{eqnarray}
with $\Delta_b = \sum_k b_k \Delta_k$, where $b$ are two-dimensional complex unit vectors and the sum $\sum_b$ runs over all neutral physical Higgses $H^0_b$. 

\section{Case $n_R=1$}
\label{3x1}

First we consider the minimal extension of the standard model by adding only one right-handed field $\nu_R$ to the three left-handed fields contained in $\nu_L$.

For this case we use the parametrization $\Delta_i = \left(\sqrt{2}\, m_D/v \right)\vec{a}_i^T$, where $\vec{a}_1^T = \bigl(0,0,1\bigr)$ and $\vec{a}_2^T = \bigl(0,1,e^{i \phi}\bigr)$. Diagonalization of the symmetric mass matrix $M_{\nu}$ (\ref{Mneutr}) in block form is 
\begin{equation}\label{Mneutr1}
U^{T}M_{\nu}U = 
U^{T}\left(\renewcommand{\arraystretch}{0.8} \begin{array}{cc} 0_{3\times3} & m_{D} \vec{a}_1\\
m_D \vec{a}_1^T & \hat{M}_R \end{array} \right)U=\left(\renewcommand{\arraystretch}{0.8} \begin{array}{cc} \hat{M}_l & 0 \\
0 & \hat{M}_h \end{array} \right).
\end{equation}
The non zero masses in $\hat{M}_l$ and $\hat{M}_h$ are determined analytically by finding eigenvalues of the hermitian matrix $M_{\nu}M^{\dagger}_{\nu}$. These eigenvalues are the squares of the masses of the neutrinos $\hat{M}_l=\text{diag}(0,0,m_l)$ and $\hat{M}_h=m_h$. Solutions $m^2_D=m_hm_l$ and $m^2_R=(m_h-m_l)^2 \approx m^2_h$ correspond to the seesaw mechanism.
\begin{figure}[H]
\begin{center}
\includegraphics[scale=0.75]{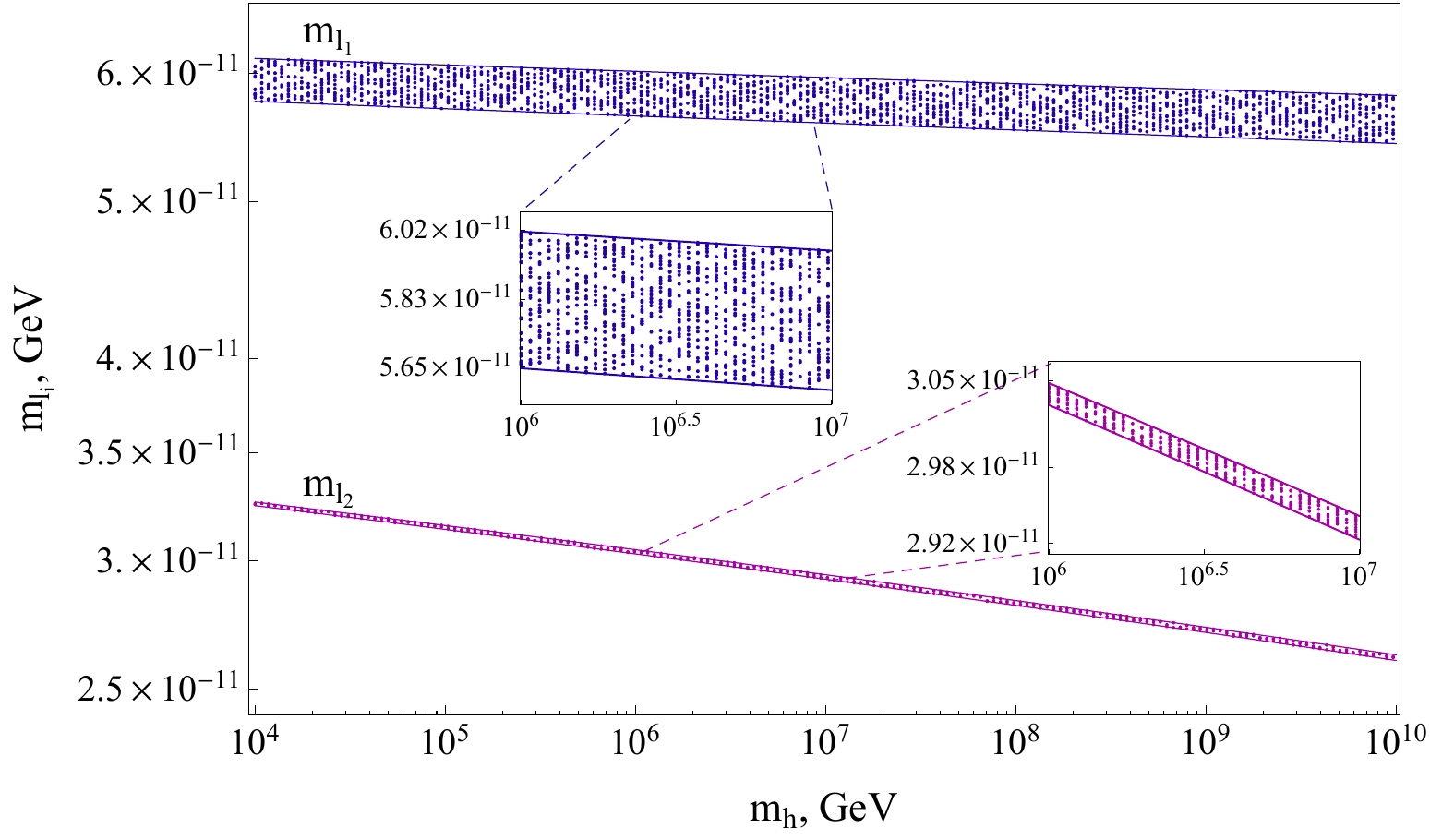}
\end{center}
\caption{Calculated masses of two light neutrinos as a function of the 
  heavy neutrino mass $m_h$. Solid lines show the boundaries of allowed 
  neutrino mass ranges when the model parameters are constrained by the
  experimental data on neutrino oscillations 
  with $\theta_{\mathrm{atm}}=45^{\circ}$. 
  The allowed values of $m_{l_1}$ and $m_{l_2}$ form bands, their scattered
  values are shown separately in the middle plots.}
\label{picture1}
\end{figure} 
The diagonalization matrix $U$ for the tree level is constructed from a rotation matrix and a diagonal matrix of phases $U_{\text{tree}}=U_{34}(\beta)U_i$, where the angle $\beta$ is determined by the masses $m_l$ and $m_h$.

For the calculation of radiative corrections we use the following set of orthogonal complex vectors: $b_Z = (i,0)$, $b_1 = (1,0)$, $b_2 = (0,i)$ and $b_3 = (0,1)$. Diagonalization of the mass matrix after calculation of one-loop corrections is performed with a unitary matrix $U_{\text{loop}}=U_{\text{egv}}U_{\varphi}(\varphi_1,\varphi_2,\varphi_3)$, where $U_{\text{egv}}$ is an eigenmatrix of $M^{(1)}_{\nu}M^{(1)\dagger}_{\nu}$ and $U_{\varphi}$ is a phase matrix. The second light neutrino obtains its mass from radiative corrections. The third light neutrino remains massless.

\begin{figure}[H]
\begin{center}
\includegraphics[scale=0.8]{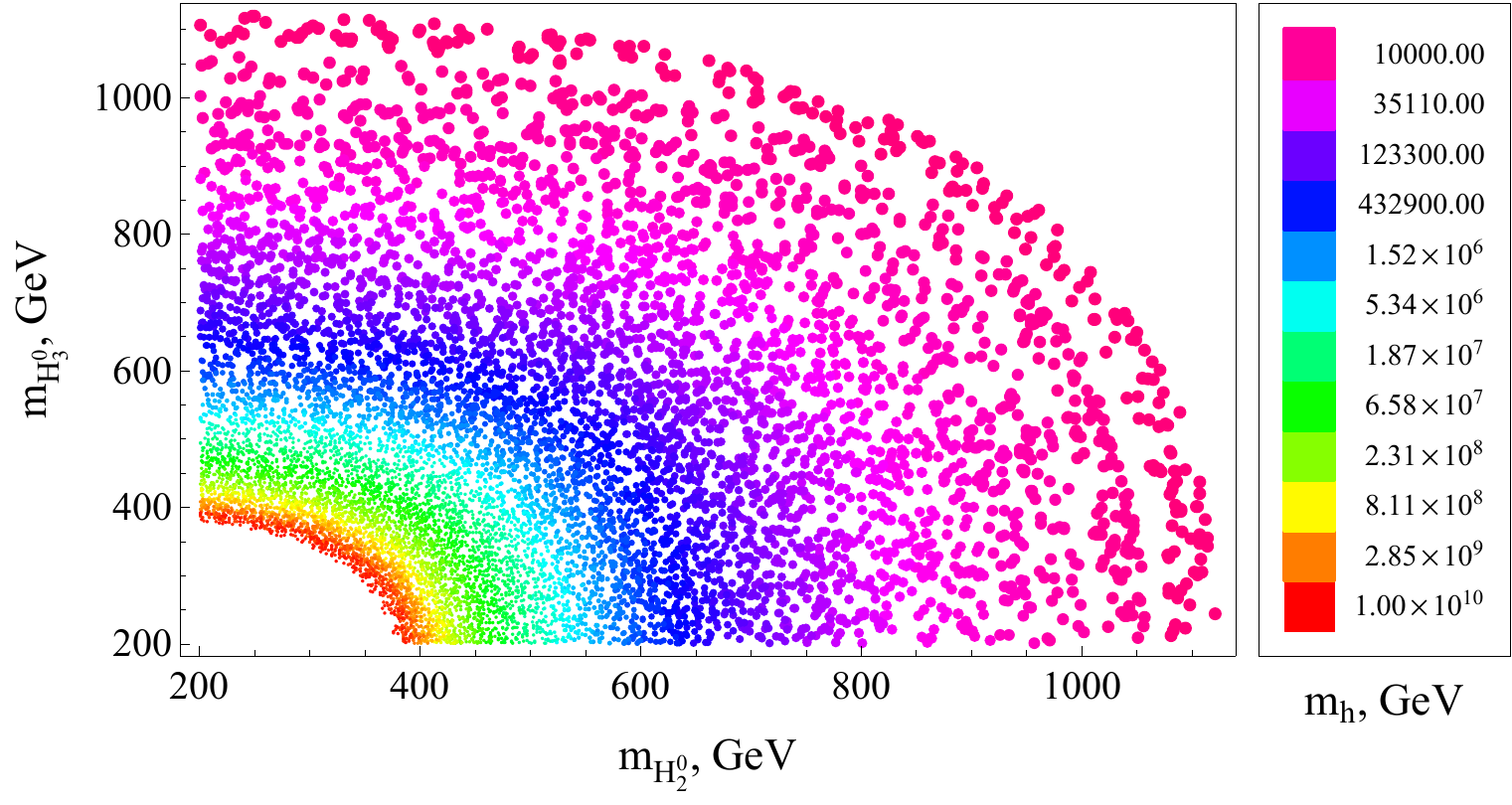}
\end{center}
\caption{The values of the free parameters $m_{H^0_2}$ and $m_{H^0_3}$ as functions of
  the heaviest right-handed neutrino mass $m_{h}$, for the case
  $n_R=1$. The mass of the SM Higgs boson is fixed to $m_{H^0_1}=125$~GeV and the angle of oscillations is $\theta_{\mathrm{atm}}=45^{\circ}$.}
\label{picture2}
\end{figure}

The masses of the neutrinos are restricted by experimental data of solar and atmospheric neutrino oscillations \cite{Gonzalez} and by cosmological observations. The numerical analysis shows that we can reach the allowed neutrino mass ranges for a heavy singlet with the mass close to $10^4$~GeV and with the angle of oscillations fixed to $\theta_{\mathrm{atm}}=45^{\circ}$, see Fig.~\ref{picture1}. The free parameters $m_{H^0_2}, m_{H^0_3}$, and $\phi$ are restricted by the parametrization used and by oscillation data. Figure
\ref{picture2} illustrates the allowed values of Higgs masses for different values of the heavy singlet.

\section{Case $n_R=2$}
\label{3x2}

When we add two singlet fields $\nu_R$ to the three left-handed fields
$\nu_L$, the radiative corrections give masses to all three light
neutrinos.

Now we parametrize $\Delta_i=\frac{\sqrt{2}}{v}\,\left(m_{D_2} \vec{a}_i^T , m_{D_1} \vec{b}_i^T  \right)^T$ with $|\vec{a}_1|=1$, $|\vec{b}_1|=1$, $|\vec{a}_2|=1$ and $|\vec{b}_2|=1$. Diagonalizing the symmetric mass matrix $M_{\nu}$ (\ref{Mneutr}) in block form we write: 
\begin{equation}\label{Mneutr2}
U^{T}M_{\nu}U = 
U^{T}\left(\renewcommand{\arraystretch}{0.6} \begin{array}{cc} 0_{3 \times 3} & m_{D_2} \vec{a} \hspace{.2cm} m_{D_1} \vec{b}\\
\begin{array}{c} m_{D_2} \vec{a}^T \\ m_{D_1} \vec{b}^T \end{array} & \hat{M}_R \end{array} \right)U=\left( \begin{array}{cc} \hat{M}_l & 0 \\
0 & \hat{M}_h \end{array} \right).
\end{equation}
The non zero masses in $\hat{M}_l$ and $\hat{M}_h$ are determined by
the seesaw mechanism: $m^2_{D_i}\approx m_{h_i}m_{l_i}$ and $m^2_{R_i}
\approx m^2_{h_i}$, $i=1,2$. Here we use $m_1>m_2>m_3$ ordering
of masses. The third light neutrino is massless at tree level.

The diagonalization matrix for tree level $U_{\text{tree}}=U_{\text{egv}}^{\mathrm{tree}}U_{\phi}(\phi_i)$ is composed of an eigenmatrix of $M_{\nu}M^{\dagger}_{\nu}$ and a diagonal phase matrix, respectively.

For calculation of radiative corrections we use the same set of orthogonal complex vectors $b_i$ as in the first case. Diagonalization of the mass matrix including the one-loop corrections is performed with a unitary matrix
$U_{\text{loop}}=U_{\text{egv}}^{\mathrm{loop}}U_{\varphi}(\varphi_i)$, where
$U_{\text{egv}}^{\mathrm{loop}}$ is the eigenmatrix of
$M^{(1)}_{\nu}M^{(1)\dagger}_{\nu}$ and $U_{\varphi}$ is a phase
matrix.

A broader description of the case $n_R=2$ and graphical illustrations of the obtained light neutrino mass spectra is given in Ref.~\cite{jurc}. Both normal and inverted neutrino mass orderings are considered.

\section{Conclusions} \label{concl}

For the case $n_R=1$ we can match the differences of the calculated light neutrino masses to $\Delta m^2_\odot$ and $\Delta m^2_\mathrm{atm}$ with the mass of a heavy singlet close to $10^4$~GeV. The parametrization used for this case and restrictions from the neutrino oscillation data limit the values of free parameters. Only normal ordering of neutrino masses is possible.

In the case $n_R=2$ we obtain three non vanishing masses of light neutrinos for normal and inverted hierarchies. The numerical analysis \cite{jurc} shows that the values of light neutrino masses (especially of the lightest mass) depend on the choice of the heavy neutrino masses. The radiative corrections generate the lightest neutrino mass and have a big impact on the second lightest neutrino mass.

\bigskip
\noindent \textbf{Acknowledgements} 

The authors thank Luis Lavoura for valuable discussions and suggestions. 
This work was supported by European Union Structural Funds project "Postdoctoral Fellowship Implementation in Lithuania".

\bigskip
\newcommand{\hepth}[1]{\href{http://arxiv.org/abs/hep-th/#1}{\tt hep-th/#1}}
\newcommand{\hepph}[1]{\href{http://arxiv.org/abs/hep-ph/#1}{\tt hep-ph/#1}}
\newcommand{\nuclth}[1]{\href{http://arxiv.org/abs/nucl-th/#1}{\tt hep-ph/#1}}
\newcommand{\arXiv}[1]{\href{http://arxiv.org/abs/#1}{\tt arXiv:#1}}

 \end{document}